\documentclass[a4paper]{article}
\usepackage[dvips]{graphicx}
\newcommand{\be}{\begin{equation}}
\newcommand{\ee}{\end{equation}}
\newcommand{\bel}[1]{\begin{equation}\label{#1}}
\newcommand{\bea}{\begin{eqnarray}}
\newcommand{\eea}{\end{eqnarray}}
\newcommand{\beal}[1]{\begin{eqnarray}\label{#1}}
\newcommand{\nn}{\nonumber}
\newcommand{\nin}{\noindent}
\newcommand{\nfn}[1]{\renewcommand{\theequation}
           {#1.{\arabic{equation}}}\setcounter{equation}{0}}
\def\d{\partial}
\def\xp{x_{\perp}}
\def\ra{\rightarrow}
\def\P{\hbox to 10pt {\hfill \large $\cal P$} }
\def\F{\hbox to 10pt {\hfill \large $\cal F$} }
\def\V{\hbox to 10pt {\hfill \large $\cal V$} }
\def\FC{\hbox to 20pt {\large $\cal FC$ } }
\def\Fd{\hbox to 18pt { {\large $\cal F$} \hss {\it~d} } }
\def\dlr{\hbox{\raisebox{8pt}{$\leftrightarrow$}$\!\!\!\!\!\d$}}

\begin{document}

\title{\bf About the structure of the Froissart limit in QCD}

\author{ {\sc O.V. Kancheli}\thanks{E-mail: kancheli@vxitep.itep.ru} \\
     {\it  Institute of Theoretical and Experimental Physics, }  \\
     {\it  B. Cheremushinskaya 25, 117 259 Moscow, Russia. } }
\date{}
\maketitle
\begin{abstract}
 The Froissart  asymptotic behavior of cross-sections
is usually considered  in a parton picture
as corresponding to the collision of
two almost black  disks filled with partons.
~In this article we mainly concentrate on the examination
of the local transparency of such F-disks.
We discuss how is it possible to guarantee the boost-invariance of
the reciprocal transparency of two such F-disks in a process of their
collision, despite the fact that the mean area of the overlapping of these
F-disks at the same impact parameter is varying with the Lorentz
frame.
We argue that on will always have such problems, if the
dominant interactions at all energies remain soft,  but such a trouble
can be probably avoided if the mean parton virtualities grow with energies.
This last is natural in QCD, and we use the qualitative generalization
of BFKL approach to estimate the distribution of hard partons with various
virtualities inside a F-disk.
As a result, the quasiclassical partonic wave function corresponding
to the F-limit can be approximately represented by the system of enclosed
parton-gluon disks with a growing virtuality and blackness.
With the increase of energy the new disks with larger virtualities are
created in the middle of the previous disks, and then they expand with
the same transverse velocity.
\end{abstract}
\newpage
\setcounter{footnote}{0}
\section*{\bf 1. Introduction.}
\nfn{1}
This section has mainly a review character. We draw here attention
to some problems connected with the dynamical structure of the
``Froissart disk" and their description in terms of different
approaches.

The Froissart (\F)
\footnote{ In this article we also use the sign \F to denote
the corresponding object(quasiparticle ?), which is sometimes called
Froissaron}
asymptotic behavior \cite{Froi} corresponds to the growth of total
cross sections like $\sigma_{tot}~\sim~Y^2$,
where~$Y=\log~s$.
This behavior is proved to be maximally fast in field theories
with massive particles.
In Regge-Glauber approach the \F behavior arises naturally after the
s-chanal unitarisation in models with the supercritical pomeron (\P )
- with the trajectory
$\alpha(t) \simeq  1 + \Delta +\alpha' t ,~~~ \Delta > 0$.~
This follows already for a simple eiconal set of diagrams~\cite{ChengWu}.~
In QCD the perturbative \P corresponds to the BFKL-like
generalization \cite{BFKL} of a  two gluon exchange, and is
very likely supercritical.
All exsisting experimental data on high energy hadron cross sections
also point on a supercritical \P.
Due to all this, it is accepted to believe that the \F behavior of
cross-sections becomes apparent when energies become essentially large.

 The supercritical \P  , when we write its contribution to an elastic
scattering amplitude in the representation of impact parameters as
\bel{pom1}
v(Y, \xp ) =    \frac{g_a g_b}{\alpha^{\prime} Y_1 } \cdot
         \exp (~ \Delta Y_1  ~-~  \xp^2 / 4 \pi \alpha' Y_1 ~) ~,
\ee
where $Y_1  = Y + i \pi/2$  ~,~~ corresponds, after an eiconal-like
unitarisation of the S-matrix, to the expressions
\beal{eic1}
S(Y, \xp ) ~=~ e^{-v} ,~~~~~~~
A(Y, \xp ) ~=~  i ( 1  - e^{- v}   ) ~~~,           \\
\sigma_{in}(Y, \xp) ~=~   1  - e^{- 2 \Re v}  ~,~~~~~
\sigma_{tot}(Y, \xp) ~=~  2 ( 1  - \Re e^{- v}   )~,  \nn
\eea
which have a very distinctive property.
For a  large $Y$ from (\ref{eic1}) it follows approximately that
\bel{teta}
 \sigma_{in}(Y, \xp) ~=~ (1 - T)~ \theta \Big(~R(Y) - \xp~\Big)~,
\ee
where  $R(Y) = Y \cdot 2\sqrt{\Delta \alpha'}$ ,
and the transparency $T(Y,\xp) \ra 0$ when
$Y \ra \infty$ and $\xp/R(Y) < 1$~.
This corresponds to an almost black disk with a radius $R(Y)$ whose
border is spread by $\delta~R(Y) \sim \sqrt{\Delta \alpha'}$.~
It expands linearly with $Y$, and thus leads to the \F type behavior
of cross-sections
~$\sigma_{tot} = 2~\sigma_{in} \simeq \pi~R^2 (Y) \sim Y^2$.  \\
The explicit eiconal dependence $S[v]=e^{-v}$ of $S$ on $v$ is
by itself not essential to reach such conclusions,
and instead of (\ref{eic1}) one can in the same way work with
generalized eiconal series
\bel{eics}
A(Y, \xp ) ~=~  i \sum_{n=1}^{\infty} \frac{c_n}{n!}(-v)^n~,
\ee
where $c_n$~ are nearly arbitrary positive coefficients, representing
the contribution of diffraction generation beams.
The general method of working with series (\ref{eics}), in a connection
with the \F behavior, was suggested in the nice work of Cardy \cite{Cardy},
and by means of that one can in a simple way take into account also more
complicated reggeon diagrams, not only a nonenhanced one (eiconal type).
Various aspects of this approach are developed further in a number
of papers
\cite{Amati1,LRfroi,LeRy1,DubTer,kl}.
The main point here is that we start from a black disk
(of type (\ref{eic1}) or (\ref{eics})) as a zero approximation,
and then find correction to it -
this changes only slightly the whole picture.
Corrections come from processes that take place on the borders of
this \F disk (\Fd) - at $\xp \simeq R(Y)$ - when the impact parameters
are such that two colliding \Fd touch each other by their spread
borders.
It corresponds to various diffraction generation and multipomeron
processes, and one must take care that their cross-sections be not
larger than the cross-sections of the main processes.
Corrections to the internal parts of \Fd cancel in this approach due
to an elastic screening \cite{Cardy},
that is in fact the screening in the process of interaction. \\
The resulting picture of \F corresponds to the soft and black \Fd, and
when interpreted in parton terms, is probably contradictory,
as we discuss in this article.

In this paper we consider the transparency of the internal parts of \Fd.
This small quantity can be more sensitive to t-unitarity constraints,
and is directly connected with the $S(Y, \xp)$-matrix, that gives the
amplitude for a target to tunnel through \Fd at a given $\xp$ without an
interaction ~-~ the corresponding transparency  ~~$T ~=~ |S|^2$.~

Let's begin with comments of the parton interpretation of some basic
regge expressions corresponding to the supercritical \P.~
In the parton language the \P contribution to $\Re v(Y, \xp)$ is
proportional to the mean transverse density of low energy partons
in a fast hadron.~
If the interactions of these partons with a small target are
independent from each other, then the mean number of such interactions
- $<n>$ is also proportional to $v(Y, \xp )$.
Because in this (uncorrelated) case the probability of some number
of interactions is given by Poison distribution,
the probability of {\em no} interaction $T$
is simply $\exp (- <n>)$,  and this is what corresponds to
the eiconal $S \sim \exp (-v)$ ,
or to the close to (\ref{eics}) forms of the $S(Y,~\xp)$-matrix.

The possible opposite situation is when the correlations, coming
from large fluctuations in the target and in the incoming \Fd,
are maximale.
For example, if the size $r_t$ of the target fluctuates to small values,
its interaction cross-section $\sigma_0(r_t)$ with all \Fd partons
becomes small.
As a result the mean number of interactions with the target in these
configurations $<n> \sim v \sigma_0(r_t)$ can be also small.
But the probability to come in this state can be larger than the
probability to ``not interact"  $e^{-2v}$ in the main configuration.
In all QCD-like field-theories such probability for the target
to fluctuate to small $r_t$, ~( in such a way to have $<n> \sim 1$),~
is of the order $\sim 1/v^2$. ~It then leads to
\bel{peic}
    S ~=~ \frac{1}{1+v}  ~, ~~~~~~~~~~A ~=~ i v/(1+v)
\ee
Or one can consider such a fluctuation (in fact the corresponding
small component of the stationary Fock function) in a fast hadron,
that doesn't contain the \Fd at all. This probability is always
$\sim\exp (-c Y)$, that leads again to the power-like $S[v]$ of type
(\ref{peic}). In terms of series (\ref{eics}) such a behavior
corresponds to a fast grow of ~$c_n~\sim~n!$~~,~  and so leads
to the $S$ matrix decreasing much slowly with $v$~
\footnote{
 Evidently for a large $v$ such series don't converge
and we must use some indirect method of summation of diagram
contribution. This also shows that the decomposition over reggeon
diagrams is not always the best way, alredy in such a simple
situation.}.
 But we still have a black disk at large $v$. \\
This example in particular shows that an elastic $S$ matrix, although
purely diffractive (the amplitude $A~\simeq$ imaginary),
can be defined at some $\xp$ not with main inelastic processes.
And it is probably a rule and not an exclusion.

The essential point is  that  $v \ra \infty,~~ S[ v ] \ra 0$
when $Y \ra \infty$
~-~ and all this needs the infinite growth of the parton density
in a bare supercritical \P with $Y$  ~-~
as a result follows the blackness of the \Fd.
The various forms of $S[v]$ in (\ref{eics}) only represent the
different patterns of screening of these partons one by another
in process of their interaction with the target, when their density
becomes large, and then the nearly black disk is used as a next
approximation.

But let take firstly into account that partons interact - shadow
one another and recombine in all rapidity interval and not only when
they interact with the target.
In the quasiclassical approximation (without \P loops), to estimate this
effect one must sum all tree diagrams with \P ~.
After that one can (possibly) put the result in eiconal-like series
(\ref{eics}), to take into account also the screening in the process of
interaction with a target.~~~
For a large \Fd one can neglect the transverse motion (in $\xp$) of \P
inside the tree diagrams.
Then the full contribution from the sum of all such diagrams can be
represented in the simple form
\bel{Vv}
   v  \ra  V(Y,\xp)
         ~\simeq~ \frac{v}{1 +  \lambda ~v} ~~,~~~~~
           \lambda ~\sim~ \frac{r_3 } { \Delta } ~,
\ee
where $r_3$ is  the 3\P vertex.
In the one-dimensional parton language the Eq.(\ref{Vv}) corresponds to
the solution of the equation ~$\d V/\d y ~=~ \Delta V - r_3  V^2$~~
for the mean parton number $V$,
where $\Delta$ gives the probability for a parton to split on an unit
interval of rapidity, and $r_3$ the probability for two partons
to recombine.

So, in this case, the amplitude $V(Y,\xp)$ and not $v$ is
proportional to the transverse parton density, and we have saturation,
corresponding to the gray disk $V \ra \lambda^{-1}$ at $Y \ra \infty$,
and not an infinite growth of the density.
Then, instead of (\ref{pom1}),(\ref{eic1}), the ``additionaly" unitarised
S-matrix and the amplitude representing the gray \Fd take the form
\beal{gr-eic}
 S_1(b,y)=  \sum_{n=0}^{\infty} \frac{c_n}{n!}(-V)^n~
 ~~~,~~~~~~~~
 F_1(b,y) ~=~ i\Big( 1 ~-~  S_1(b,y) \Big)~~.~~
\eea
Here we can face with some problems, because higher order diagrams
with \F loops (containing $V$ or $F_1$ as propagators,
like in (\ref{gr-eic})), can give arbitrary larger powers of
$Y$ in amplitudes.
And then one must again consider the full series of such diagrams
and hope that it will converge to a sensible gray \F limit
(or now a black one~?), or consider all this as an evidence that
the gray disk picture is inconsistent.

But such arguments are slightly naive.
Firstly because the series of diagrams with \F as a quasiparticle
are not the good ones, despite the fact that the series
of all reggeon diagrams with \P can be formally represented
\cite{Cardy}as new series over the \F like objects of
type~(\ref{eics}).~
It is because the initial(perturbatie) vacuum of a supercritical
\P is an unstable one,
and the ``quasiparticle" \F represents in fact the growth with
~``time" ~$y$~ of a bubble of the other (stable) phase of the \P
system. If we write \P Lagrangian in the form
\bel{rftl}
 {\cal L} ~=~ \psi^+ \frac{~\dlr }{2~\d y}\psi  ~+~
 \Delta \psi^+ \psi ~+~
\alpha' ~\vec{\d_{\perp}} \psi^+  ~\vec{\d_{\perp}} \psi  ~+~
r_3 ~(\psi^+\psi^+\psi + \psi^+ \psi\psi) + ~...~ +
(\psi^+J + J^+\psi)~,
\ee
where $\psi(y,\xp)$ is \P field and $J$ - sources, representing
``external" hadrons, then it is evident that at $\Delta >0$
the vacuum $\psi =0$~ is unstable and the classically stable vacuum
can be at $\psi = \Delta /r_3$.~
For the nonrelativistic system, described by (\ref{rftl}), the initial
vacuum is in $\psi =0$ state, and can not by itself tunnel to the other
phase.~ But an external particle $J$ can inject \P's in the vacuum and
then it leads to the growing bubble of other stable phase.
As a result, in the diagrams with \F loops the different \F's
(in fact ``parallel condensats of $\psi$ ") partially occupy the same
place in $\xp$ plane during the ``time" $y \sim Y$, with the area of
$y*\xp$ overlapping ~${\cal S} \sim Y^3$.
Then one must probably add explicitly (or effectively - as a result of
summation of some series of \F subdiagrams) to every such a diagram
a factor $\exp(- {\cal S})$ representing the probability that
the ``parallel" ~\F's don't interact (like the Sudakov factor).~
And in this case the contributions of all diagrams with \F loops
would not lead to higher powers of $Y$, and can be essentiall
only on the border of \Fd when ${\cal S}$ is small.
This shows that the consistency of a gray picture of \Fd must be
probably checked with another methods.   \\
It should be noted, that the gray \Fd can be acceptable
as an approximation (and possibly in a very broad energy interval )
~\cite{CaDe}, before the hard component of \Fd becomes large.

It is known scene then, the strong coupling regime with
the asymptotic behavior of cross sections of type
~$\sigma_{in} \sim Y^a ~,~~~ 0  \leq a \leq 2$~
has been introduced \cite{GrMig},
that in the limiting cases $a=0 ~,~a=2$ the infinite number of
identities  between vertices must be fulfilled, that is very unnatural,
because it needs fine tuning of the infinite number of constants
(bare vertices), coming from large distances.
Despite that, let briefly mention the peculiar properties of such a \F.
If we choose the \F Green function and the 3\F vertices in the scaling
form
\beal{fgf}
 ~~~~~G_{\omega} (k) ~=~ \omega^{-3} \phi (k/\omega)~~,~~~~~~~~  \nn \\
 \Gamma_3 (\omega,\omega_1,\omega_2, k_i) = \omega^3 \gamma (k_i/\omega)~,
\eea
(~where $\omega = j-1$~)
then the Dyson equation for $G$ and $\Gamma_3$ can be fulfilled
on the level of their singular parts.
The same is true for higher \F vertices  $\Gamma_n$.
For example, the $3\Gamma_3$ term in the Dyson equation for $\Gamma_3$
reduces like
\be
  \int d\omega ~d^2k ~G^3 ~\Gamma^3_3  \sim \omega^3 \sim  \Gamma_3
\ee
All this also corresponds to the conditions, that can be described
in a simbolical form as
\bel{ward}
     G \Gamma_n  ~=~  1  ~~~~,
\ee
and that leads to a cancelation of all ``superfluous" poles in
enhanced \F diagrams, and  therefore they are effectively reduced to
unenhanced one.
As a result the total cross section is given by the sum of multi-\F
exchange diagrams with all terms of the same order in $Y$,
and one can hope that the rate of convergence of these series
doesn't depend asymptotically on $Y$ itself. \\
Because the corresponding \Fd is gray, the 3\F diffraction
(of states with high masses) should be expected high
($\sim \sigma_{tot}$).
And now it comes not only from the border of \Fd.
Using the Eq.(\ref{fgf}) one can estimate that
\be
    \d \sigma_{dif} /\d \eta \sim \eta  ,
\ee
where  ~~$\eta = \ln M_{beam}$.
So one can expect that for such a \F there are large fluctuations
in individual events at  the same impact parameter.
Thus such a gray \Fd corresponds to a picture of a growing disk (bubble)
filled by pomerons (or, in the other language, by corresponding partons),
that are itself in a critical point state~
and not in a new stable phase like $\psi \sim\Delta /r_3$.

If we want to make transition from \P to \F using reggeon
diagrams the main trouble is the following one.
Reggeon diagrams can be safely applied only when the mean
transverse density of reggeons is not a large one - pomerons should
not cover one another, because they are purely composite objects
and can completely dissolve, when their density exceeds some
critical value.~
This condition leads \cite{aggk} to the ``limiting" order
~$n_{max} \sim Y$~ in all series over the reggeon diagrams,
like (\ref{eics}).~~
But in all expressions (\ref{eic1}),(\ref{eics}) the mean essential
$n \sim \exp Y\Delta$ is much higher then such $n_{max}$ at
$Y \gg 1/\Delta$,~
and so the prepared \F takes in fact
the main contribution from the values of $n$ located outside
the region of the \P 's applicability.
And, as a result, if we cut series on the order ~$n \sim Y$,~
different methods of summation can  give different answers. \\
The situation may be slightly better when the effective transverse
size of \P 's decreases, when their density becomes so large that
\P 's begin to touch one another, so that as a result they don't
dissolve.~
Or, in other words,  more and more hard \P 's become essential ,
whose size (and correspondingly the virtuality) is defined by their
total number depending on $Y$.
If \F is constructed out of such objects, the structure of the internal
part of \Fd can be nonstationary and becomes more and more black with
the growth of energy.

In this paper we discuss these troubles with a \Fd structure
directely in a parton approach
(started and mainly developed by Levin and Ryskin~\cite{LRfroi}).
It has probably a much higher region of applicability,
because we don't introduce here auxiliary quasiparticles
(like \P) and work directly with partons (gluons).~
But this way is  much more complicated, especially for
the unification with the existing high energy phenomenology.

It is interesting that in the parton language we meet
the close problems with the structure of \Fd.
When we restrict ourselves with only soft partons it is too difficult
to produce the black \Fd, and this leads to troubles, that we discuss
from a slightly different point of view in section~2~.

In next sections we try to discuss a part of these problems.
Because we expect that for \Fd the dense murtiparticle states
are most essential we will not consider directely the Fock space
vectors for partons, but only the corresponding qusiclassical
densities and probabilities.
 \\
\noindent    The outline of this paper is the following. \\
\hspace*{10pt}
In {\bf Section 2}  the general restrictions on the structure
of \Fd  are considered, that follow from the requirement
of the longitudinal Lorentz (boost) invariance (frame independence)
of calculations of the ~(\Fd x \Fd) collisions cross-sections. \\
\hspace*{10pt}
In {\bf Section 3} we describe the BFKL inspired parton model for \Fd
giving the distributions of hard partons with various virtualities
in  \Fd  and their variation with $Y$. \\
\hspace*{10pt}
In {\bf Section 4} we consider the (\Fd x \Fd) collision in such
a hard parton model and estimate the transparency and their
dependencies  on the longitudinal  Lorentz system. \\
\hspace*{10pt}
In {\bf Section 5} we consider a regge model where such a hard  \F
behavior can arise and where the hard part \P is included in BFKL
like manner.  \\
\hspace*{10pt}
In {\bf Section 6} we mention a number of questions related
to the \F type behavior - like of the possible appearance of the \F
embryo in existing experimental data;~~
to the similarity of \F x\F  collisions with the heavy AxA
collisions ;~~
and to the question - if there is any limit from above on the \F -like
behavior, possibility connected with gravitational interactions.
%
\section*{\bf 2.   Unitarity in t-channel,
      longitudinal Lorentz invariance and  the \F ~limit }
\vspace{1pt}

The general limitations on the structure of  the \F behavior
coming from the S-channel unitarity are in some sense trivial
and direct - one must ``only" carefully calculate all parton currents
and exclude multiple counting of cross-sections of the same processes.
In particular, the cross-sections of subprocesses must be smaller then
the cross-sections of the main processes like
$\sigma (\hbox{\small \it diffraction generation}) <  \sigma_{tot}$~.

The possible restrictions on the \F behavior coming from the
t-unitarity are much more complicated, because one must continue
these conditions from t-channel, and this is a nontrivial problem
in the parton picture.
For \F type behavior the most essential are the n-particle states
in t-channel with $ n \ra \infty $ when  $ s \ra \infty $ .
And it is in fact unknown how to take the unitarity restrictions
on them into account.
In the reggeon field theory(RFT) all t-unitarity conditions are
automatically fulfilled, when we sum all diagrams.
But at high reggeon density the RFT is unapplicable.
And in the parton picture, when we  directly consider an interaction
of stationary hadron Fock states and then have no limitations on
the value of parton density, there is no direct way how to control
possible restrictions coming from the t-unitarity.

Here  we use some approximate conditions which can be partially
equivalent to the mean form of the t-unitarity for
multiparticle amplitudes.
It was proposed in \cite{aggk}, that the t-unitarity restrictions
would be equivalent on average to the longitudinal Lorentz (boost)
invariance of all cross-sections, calculated in a parton picture.
So if we calculate some cross-section using the partonic wave functions
$\psi (p)$ and  $\psi (p_{b})$ of fast colliding hadrons with momenta
$p_{a},$ $p_{b}$ \ then  we expect that  this cross-section must be the
same in all longitudinal Lorentz frames - that is if we calculate the
cross-sections
using \ $\psi (L(\vartheta )p_{a})$ and
$\psi (L^{-1}(\vartheta )p_{b})$, where  $L(\vartheta )$
is a longitudinal boost.
Remember, that in a parton picture such boosts $L(\vartheta )$ act on
hadrons Fock state very nontrivial, changing the number of wee
partons, etc.
Probably such conditions should be imposed only on the dominant in
$1/Y$~  (or, maybe, in $\exp{(-Y)}$ ?) terms in cross-sections,
because for the nonrelativistic interaction the restrictions
from the t-channel must be absent.

No strong arguments for such general propositions are known to
the author.
But, firstly, it is absolutely natural by itself in the parton
picture that the calculations of cross-sections, if correct,
must give a frame independent answer.
And secondly, such a proposition can also be confirmed in RFT
if we give the partonic interpretation to reggeon diagrams,
by t-cutting them at various intermediate rapidities,
as if we calculate various multiparticle inclusive cross-sections.
Here we only mention an idea of the construction.

Consider all reggeon diagrams  giving a contribution to a total
cross-section. If we fix a longitudinal Lorentz frame, such that
colliding particles have rapidities $y$ and $Y-y$ - then the cutting
of all diagrams at this  $y$  can be interpreted as a calculation
of the total cross-section (and also various inclusive cross-sections)
in a parton picture - and different diagrams in Fig.1 give various
contributions, including screenings, to this cross-section.
\begin{figure}[h]    \centering
\begin{picture}(70,150)(50,60)
\includegraphics[scale=.3]{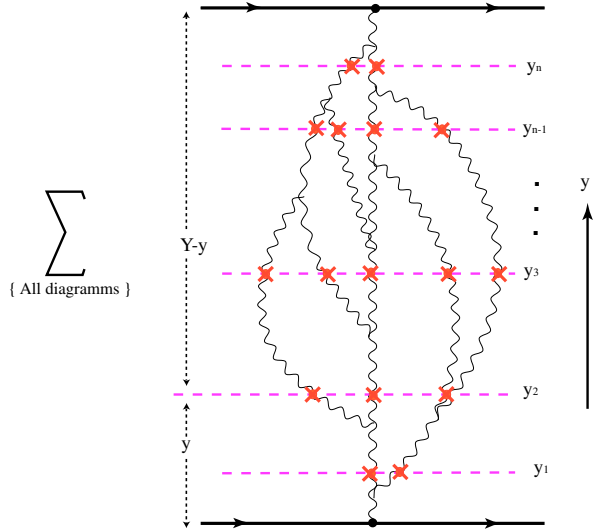}
\end{picture}
\vspace{20mm}
\caption{The t-chanel sections $y_i$ of complicated reggeon diagrams.
From one side they can be considered as representing the calculation
of the cross-sections, using the parton wave functions in various
longitudinal frames $y_i$.~
From another side - as a sequense of events in the t-channel evolution
of a pomeron system along the  ``time"~$y$ }
\end{figure}
When we go from one section to another $y \ra y + \vartheta$,~
then their partonic contributions to $\sigma_{tot}$~
(and other cross-sections)~ coming from various diagrams change,
but their sum must be the same.
The transition between  these t-intermediate sequences of the states
can be reached by a longitudinal boost.
So we move through a sequence of the intermediate states in t-channel,
and expect the self-consistency of the answers.
Evidently, it is the t-unitarity that guaranties such a consistency.

Let's apply this requirement of the frame independence to the
calculation of total inelastic cross-section in the parton approach
\footnote{In fact it is known for a long time that in a parton model
we can meet some problems with the longitudinal frame dependence.
This happens also when we try to interpret in parton terms most
pomeron models, that are by their construction frame independent,
like the week coupling pomeron \cite{wpom},
with their asymptotic universality of total cross-sections.
Similar phenomena appear in a collision of two dense (color-dipole)
chains, corresponding to the parton representation of BFKL near the
saturation limit \cite{mulsal}.
Probably in all such cases the frame independence can only be restored
in a rather complicated way, when there is a compensation between
processes with very far configurations, as is seen from s-chanel.
Or this is at all impossible  and signals about a violation of the
t-unitarity. }.~
Firstly consider the collision of two partonic clouds that are in
a state of a rare gas. This is the case normally described by the
reggeon diagrams, that, by their construction, include t-unitarity
requirements, so here we probably must not meet any problems.
Let the mean number of partons in colliding hadrons be $n(y)$,
$n(Y-y)$~; and the mean transverse radii of regions occupied by
these partons are~ $R(y)$, $R(Y-y)$ respectively.
Then the total inelastic cross-section can be represented~as:
\bel{st1}
  \sigma_{in} (Y) = \sigma_0~ n(y)~  n(Y-y) ~-~
a_2  ~\sigma_{0}^{2} ~n^{2}(y) ~n^2(Y-y) /
\Bigl(~ R^2 (y) + R^2 (Y-y)  ~\Bigr) ~+...~~,
\ee
where $\sigma_0$  is the parton-parton cross-section, ~$a_2 \sim 1$.~
The first term in (\ref{st1}) corresponds to a collision of at list
one pair of partons. The next terms describe  corrections from
screening and multipole collisions.~
For rare parton gas one can in first approximation neglect multiple
collisions and screening, that is to leave only the first term
in~(\ref{st1}). Then, from the requirement of the independence of
~~$\sigma _{0} n(y) n(Y-y)$~~ on ~$y$~ it follows the unique solution
for  ~$n(y)=n_{0}\exp(\Delta_0 y)$~ with some constants
$n_{0}$, $\Delta_0$.~
It corresponds to a pole in the complex angular momentum plane
(and not a cut!) - this  condition usually follows in
a relativistic Regge approach only from the t-unitarity.~
Moreover, if we write  the cross-section in  Eq.(\ref{st1}) with definite
impact parameter $x_{\perp}$,~ then from the frame independence of the
$\sigma_{in} (Y,x_{\perp})$ the function  $n(y,x_{\perp})$ is almost
completely fixed at $y \ra \infty$~ (See Eq.(\ref{ro1})).

Now let us consider the opposite limiting case of colliding parton
clouds, when the parton density is very high and partons fill
a transverse disk with the radius  $R(y)$. Then the total inelastic
cross-section can be determined from purely geometrical conditions -
it is defined by the area of an impact parameter space, corresponding
to the overlapping of the colliding disks :
\bel{st2}
\sigma_{in} (Y) ~=~ (1-T)\cdot \pi~ \Big( R(y)+R(Y-y) \Big)^2
\ee
Here~ $T \ll 1$~  is the local transparency of the disks that in
general can depend on  $y$, $Y-y$.~ Now, if  ~$T=const$(y) or
can be neglected,~ from the condition  of the independence of
the right hand side of Eq.(\ref{st2}) from $y$ it evidently
follows the unique solution for ~$R(y)=a\cdot y+b$~.~
So in this case we immediately come directly to  the \F behavior
of cross-sections. \\
\noindent There can be two main types of corrections to the
$\sigma \sim Y^{2}$ asymptotic.

a)  Corrections to $\sigma_{in}$ resulting from interactions on the
spread borders of the disks. Asymptotically at  $Y\ra \infty $ these
corrections are of the order $\sim Y$.

b)  Corrections coming from the refining of the value of the
transparency $T$, because in general one can expect that the
disk is gray, $T \ne 0$ and varies with $y$.
These corrections can be $\sim Y^{2}~\delta T$ in general.

Border type corrections of type a) are connected to the
diffraction generation. In this article we will not discuss them
and concentrate on processes of type b) taking place in the interior
parts of  the colliding disks when impact parameter
$B~\le~R(y)+R(Y-y)$~. \\

If the parton structure of the \Fd at $Y\ra \infty $ is mainly
generated by the soft processes (probably mostly nonperturbative),
and if this \Fd has a finite (not a growing with $Y$) longitudinal
thickness, then it is natural to expect that the \Fd is gray (and not
a black one).
That is the transverse local parton density inside the \Fd
is asymptoticaly finite and doesn't change with $Y$, and as a result
the value of the soft transparency $T$, entering (\ref{st2}), doesn't
decrease. This last is especially evident when we move to the
laboratory frame of one of the colliding particles.
In this case $y\sim 1$, and the fast disk with the radius $R(Y)$
collides with a standing hadron, containing now one-two partons.
And then we must expect that there is  finite probability that the
target hadron can tunnel through such a \Fd  without interaction
(this probability is a definition of the transparency~$T$).

But then we come to a contradictory situation, because in the
center-of-mass frame (c.m.) one can expect that the transparency
of two disks at the same $B$ can be much less -  for example of
the order $\exp (-Y^{2})$ - because now (on average) more partons
interact
\footnote{A problem of the same type can take place already in 2
dimensions, where $D_{\perp}=0$, if parameters describing the parton
fusion (like $r_3$) are taken very small - they are external parameters
for reggeon diagrams. In this case in the lab.frame we have $\sim 1$
parton collision with a parton cloud of another hadron, and in the
c.m.frame $\sim (\Delta/r_3) * (\Delta/r_3)$parton collisions.
For $D_{\perp}=0$ ~BFKL color dipol chains there is a similar phenomenon
\cite{mulsal}.}.
To discuss this question more carefully one should consider the possible
variation of $T$ with  $Y$ and take into account all essential parton
configurations, corresponding to the \Fd,  and also these ones that are
very far from the mean one.

In general the transparency in a high energy interaction of particles
$a$ and $b$ can be expressed as
\bel{gtr}
   T ~=~ \sum_{i,j} w^{(a)}_i w^{(b)}_j ~\tau_{ij} ~~~,
\ee
where we sum over all parton configurations of  $a$ and $b$~.
In (\ref{gtr}) $w^{(a)}_i$  and  $w^{(b)}_j$ are the probabilities
of these configurations, and  $\tau_{ij}$ - the corresponding
transparency in a ~$|i> * ~|j>$  colliding state.
One can expect that for given many parton configurations $i~j$
these transparencies are Poisson-like
$\tau_{ij} \sim  \exp{(-c N_{ij})}$,~  where $N_{ij}$ is the mean
number of parton collisions in a $~|i> * ~|j>$ scattering.
The states with the maximum parton amplitude (probabilities $w_i$)
in (\ref{gtr}) contain \Fd, but the corresponding
~$\tau_{ij}$ can be too small, and the main contribution in $T$ can
originate from a configurations very far from \Fd.
But firstly we will not take into account such a rare configuration
and consider only that being close to the mean one.

Start with  the simplest model of  \Fd :  it has the radius
$R(y)=a\cdot y+b$ and it is filled  with partons with the
saturated transverse density $\rho$ that is $const(y)$~~ (in QCD one
can expect that $\rho \sim \Lambda^4 \cdot f_{\rho}( (1/ \alpha_c)$)
inside the disk, and changes to zero on the border.
The saturated parton system with high density behaves locally (in $\xp$)
like a liquid. And the fluctuation of the density of partons is low,
because the derivations from the mean value are locally compensated
by the partons splinting or fusion.
Consider a collision of disks at transverse distances $B \le R(y)$, and
in longitudinally boosted Lorentz system in that disks have rapidities
$y,~~ Y-y$ correspondingly.
In this case the transparency  $T(Y,b)$ concedes with a probability
for two disks to go one through another without interaction.
If we take into account only the pair interactions between partons
with the cross-section $\sigma_0$, then it is simple to estimate this
probability:
\bel{tr1}
T \sim \exp ( -\sigma_0 \cdot S_{12} \cdot \rho^2_{\perp})  ~~,
\ee
where $S_{12}(Y,y,B)$ is the transverse area of disks intersections.
The area $S_{12}$ variates with $y$  and  $B$ .
It is $\sim 1$ in lab. frame, and $S_{12}\sim Y^{2}$ in c.m.frame -
and as result $T$ is also strongly dependent from $y$.
\begin{figure}[h]     \centering
\begin{picture}(150,130)(40,30)
\includegraphics[scale=.4]{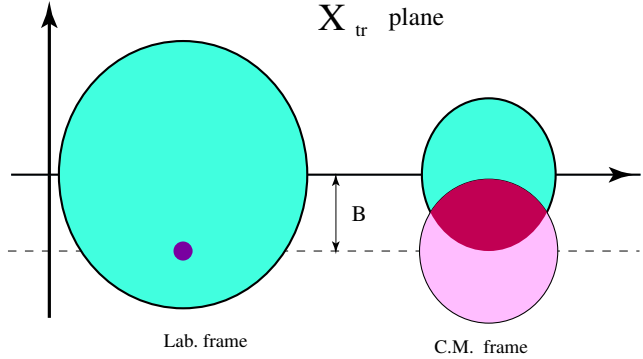}
\end{picture}
\vspace{15mm}
\caption{Two \Fd intersection in various Lorentz frames at same $B$.
Laboratory (Lab) frame and the frame close to the
Center-of-mass frame of a colliding particles (CM) are shown. }
\end{figure}
It is evident why it happens. In lab.frame only one-two partons
must penetrate through the disk, and in the c.m.frame all partons
in area $S_{ab}$ from one disk must independently penetrate through
the same layer of partons from another disk  - the last probability
is exponentially small in comparison with the first one.
The simplifications that we made during this consideration
like filling disk with parton gas and the pair interactions
between partons are not very essential for this general conclusion,
because all strong but short range corrections will probably only
renormalize parameters entering (\ref{tr1}),
and don't change $y$ dependence of $T$.~
For interaction of two gray \Fd, if the mean configurations
dominate (in $T$), one can only expect, as a generalization of
(\ref{tr1}), the dependence:
\be
T \sim \exp ( -c_{0} \cdot f_{1}(Y,y,B))~,
\ee
where \ $f_1(Y,y,B)$ is the inclusive spectrum of primary
partons at definite $B$.

\noindent What explanations for this situation can be presented,~
and how can we avoid such a trouble~?
\begin{itemize}
\item
It can be  considered as a symptom that one must leave the \F regime
as a too restrictive one for realistic field theories with finite
range interactions,
like a case with asymptotically constant cross-sections
(week coupling~~\cite{wpom}).
They are very similar in many respects.
And then the real asymptotics can probably correspond
to a strong coupling regime with $\sigma_{in} \sim Y^c  ,~~ 0 < c < 2$.
But here we will not discuss  this possibility further on.

\vspace{1pt}
\item
Suppose that all \Fd are gray with the finite and $Y$-independent
transparency in the lab. system.
But at the same time suppose also that these partons are so finely
correlated in the whole \Fd, that the same finite transparency
remains somehow also in other frames.
Then, if such  strongly correlated configurations dominate, the
Eq.\ref{tr1}  becomes unapplicable.
One cannot simply exclude this variant in general,
but it remains unclear how such a \Fd can be constructed
from massive partons with a finite range of interactions.
As an example for this gray \Fd we can mention the critical \Fd,
discussed in the Introduction.
It is hard to believe that such versions can be considered seriously
without some additional reasons.

\vspace{1pt}
\item
Search the rare parton configurations that contain a relatively small
number of partons and so can give the maximal contribution to the
transparency. Such configurations can arise from big fluctuations
in the initial state of the parton cascade.
To increase the transparency in frames close to c.m. one can ask for
an additional parton component $|\varphi >$ for a fast hadron that
doesn't contain \Fd at all and interacts slowly (or  doesn't
interact at all).~
Schematically~:
~$\psi(p_a \ra \infty) \simeq a_f  |disk~> +~ a_i|~\varphi~>$~,
~where $a_i$ is the amplitude of the $|\varphi >$ state.
It corresponds to the limiting case of Eq.(\ref{gtr}),
where we have kept only two states.
The probability for a hadron not to have a \Fd (or to have it with a
small radius less then $B$),~
is connected with $a_i$ like  ~$w(y)\sim |a_i|^2$.
In this case the expression for the transparency can be
generalized to~:
\beal{trimp}
T ~~\sim~~ \exp(-c_0 \cdot f_1 \big( Y,y,B) \big) \cdot
                    \big(1-w(y)\big) \cdot \big(1-w(Y-y)\big) + \\
~~~~~~~~~~~\tau_{\varphi d} \cdot \big( w(y) + w(Y-y) \big) + \nn \\
 ~~~~~~~~~~\tau_{\varphi \varphi } \cdot w(Y-y)) \cdot w(y)~, \nn
\eea
where for soft gray \F transparencies
$\tau_{\varphi d}$ and $\tau_{\varphi \varphi }$
are finite and $y$ independent.
Then at first sight one can expect that the last term in (\ref{trimp}),
coming from the ~$|\varphi> \cdot ~|\varphi>$~ component, can dominate
and so can make $T$  boost invariant.
But it is possible only if $w(y)$ is approximately constant for high~$y$.
And, at the same time, there are no known mechanisms that can give such
a constant $w$.~
Various estimates of $w(y)$ lead to a decreasing function of the type
$w(y)\sim \exp (-\gamma \cdot y)$.~ It coresponds to the choice at every
rapidity stage of such an evolution direction, that doesn't increase
the parton number. Such $w$  leads to the expression
\bel{trgray}
  T(Y,y) ~\sim~  \tau_{\varphi d} \cdot \big(~
      e^{-\gamma~ (Y- y)} ~+~ e^{-\gamma~ y} ~\big)~,
\ee
corresponding to the collision of the rare state  ~$|~ \varphi >$
with $|\Fd>$, and such $T$ is $y$ dependent.
Therefor on this way the soft \Fd can also not be cured.

\vspace{1pt}
\item
And finally one can consider such models of \F that the full parton
density $\rho_{\perp}$ inside the disk doesn't reach saturation at
a given $B$, but continues to grow with $y$ and varies with $B$.
In this case we can hope to compensate the $y$ dependence of
$S_{12}$ in Eq.(\ref{tr1}),~ or make the transparency of \Fd for one
parton (in lab.frame) so small (\Fd becomes more and more black with $y$)
that the configurations  ~$|~ \varphi> \cdot ~|~ \varphi>$~ and not
~$|\Fd>  \cdot ~|~ \varphi>$~ become dominant in $T$.

\end{itemize}

\nin  But firstly consider what happens  with $T$ in the parton
approach corresponding to \F in the eiconal model with the
supercritical \P, mentioned in the Introduction.
Such a \P corresponds to a chain of splitting partons,
with the simplest regge type transverse distribution
$\rho_{\perp}(y,B)  \sim y^{-1}\cdot
\exp (\Delta~ y  - B^{2}/4\pi \alpha^{\prime}y)$~.
This density \ $\rho_{\perp}(y,b)$  coinsides with the Green function
of \P. The multiple \P exchange takes into account the screening in
the process of an interaction of two hadrons so that the probability
of the interaction at given $B$ doesn't exceed $1$.
It is essential that because in this approximation 3\P vertex
(and also higher vertices) is set to zero, partons don't glue
despite their density exponentially grows.
We must here slightly refine the formula (\ref{tr1}),
because now the density depends on the transverse position :
\bel{tr3}
T \sim \exp \Big(~ -\sigma_0 \cdot \int d^{2}b \cdot \rho_{\perp} (y,b)
    \cdot \rho_{\perp}(Y-y,B-b) ~\Big)
\ee
Then we see that the integral in the exponent in (\ref{tr3})  is
$\sigma_0 \cdot \rho_{\perp} (Y,b)$ and doesn't depend on $y$,~
and as a result the contribution to the transparency also from the
\Fd configuration is boost-invariant.
The infinite growth of the soft parton density was here
essential to go to the consistent \F type behavior.

Note that here, for rare configurations, that are far from
the main one, we also have a frame independent behavior of $T$.
The maximum of $T$ is reached here when both colliding parton clouds
fluctuate to rare $\sim$(one soft parton) states. And it gives
\bel{trmul}
  T(Y,y) ~\sim~  e^{-\gamma ~y} \cdot e^{-\gamma ~(Y- y)}
            \sim  e^{-\gamma ~Y}  ~,
\ee
that corresponds to the last term in Eq.(\ref{trimp})~
The difference between (\ref{trmul}) and the gray disk case
(\ref{trgray}) is the following: here, to minimize $T$, we have
a symmetrical fluctuation, where, for a gray disk case, the maximum of
$T$ is reached on an unsymmetrical configuration.

This example raises a natural question.
Can one somehow generalize this simple eiconal mechanism to come
to \Fd with a finite density and a frame invariant $T$~?~
To understand this, let us try to find the general
$\rho_{\perp}(y,b)$ for that
the integral in (\ref{tr3}) is $y$-independent, and $\sigma_0$
doesn't depend on $y$ and $B$.
The solution of the corresponding equation for $\rho_{\perp}$ can
be represented as:
\bel{ro1}
\rho_{\perp}(y,b) \sim \int d^{2} k \cdot f_{1}(k) \cdot
                      \exp (ikb+yf_{2}(k))~,
\ee
where $f_1$, $f_2$ are arbitrary functions of $k$.
For $y \ra \infty $ the integral in (\ref{ro1}) can be taken by the
steepest decent method, so that only the neighborhoods of zeros of
$\d f_2(k)/\d k$ are essential.
Then from the positivity of $\rho_{\perp}$ it follows that $f_2$
is positive and so the dominant contribution must come from the region
$k \sim 0$, otherwise $\rho_{\perp}(y,b)$ will oscillate in $b$.
So in the essential region $f_2(k) \simeq c_1 -c_2 k^2$, and
in fact we return to the pole-like form of \P~~
\footnote{
For a collision at large $B \simeq a Y$, that corresponds
to the collision of \Fd's with their borders, the first line in
(\ref{trimp}) becomes dominant and gives $T \sim 1$.
Here to realize the frame independence of $T$ on must carefully
adjust the soft $\rho_{\perp} (y,b)$ and the dependence of the \Fd 's
radius on $Y$, to slightly separate the borders of two colliding
\Fd 's in c.m. frame in comparison to the lab.frame.
The nonlinear dependence of $R(Y) \sim a Y - c\log{Y}$ helps in
the case of dense \Fd with the sharp border.
It can be considered as coming from the
surface tension on the border of the growing new phase bubble:
$\d R /\d y \sim const - 1/R$. \\
But if the border of \Fd is diffuse, then the parton density
$\rho_{\perp} (y,b)$ must perhaps be close to Gauss form,
like in (\ref{ro1}). \\
This note concerns the most models of \Fd,
because the \Fd's border is always expected to be soft and 'gray'.}.

Here we also describe briefly mechanisms, responsible for the
transparency of the gray \Fd in the regime of the strong coupling
mentioned in the Introduction.~
For such a \Fd all diffractive and other processes with a high
derivation of density from the mean value are of the same order in $Y$
as a $\sigma_{tot}~\sim~Y^2$.
So fluctuations in individual events at $Y \ra \infty$ must be
very high.
For example, the inclusive spectra for rapidity gaps of length $y_1$
can be found from cutting the self energy diagram of the type
$\big[ ~G~ (\Gamma_3 G G \Gamma_3 ) ~G~ \big]$, where $G$ and $\Gamma_3$
are given by (\ref{fgf}).  From here the mean multiplicity of such
gaps of length $> y_1$ can be estimated as:
$$
     n_g(>y_1)  ~\sim~ Y/y_1
$$
For $y_1 \sim Y$ this in fact gives the value of the
cross-section of dif.generation ($\sim \sigma_{tot}$),
and for small $y_1$ we have $n_g(y_1 \sim 1) ~\sim~ Y$. This shows
that the average parton state looks like a gas(liquid ?) in
the critical point, with a high fluctuation of the relative density
of the order $\sim 1$ on a scale of the full system size. \\
It can then explain how the finite transparency of two \Fd 's in the
center-of-mass frame can be achieved.
Due to large fluctuations in such a parton system there is an
$y$-independent probability of order $\sim 1$ that the
soft partons of one \Fd collide with a state of other \Fd~,~
containing a large rapidity hall (gap) with no soft partons
in the same interval of rapidity. Then disks can freely move
one through another. \\
But such a critical \Fd construction needs a fine tuning of an
infinite number of parameters and looks too artificial -
we will not consider it further.

\noindent So in fact there remain two asymptoticaly different
possibilities  :

\vspace{1pt}
1) If \Fd consists only of soft partons, their transverse density
must grow with $Y$, for example like  $\sim \exp{\Delta y}$ or slower.
In terms of RFT with the ${\cal L}$ given by the Eq.(\ref{rftl})
it corresponds to such a parameter choise, that the $\psi$ system
has no ground state and $\psi$ continues to grow with the 'time' $y$,
inside the \F bubble.
If the longitudinal size $L_z(Y)$ of this \Fd, where the partons with
lowest momenta are distributed, is finite $\sim 1/m$ and doesn't grow
with $y$ asymptotically (as it is usually believed),
then this version must be probably also abandoned.
But if $L_z(Y)$  grows with $Y$, so that at least the 3-dimensional
density of soft partons remains constant, the situation can change.
In fact, already the very slow growth of
~$\rho_{\perp}(y,b) \sim \log^{1+a} y,~a>0$~ is quite enough to make the
contribution from ~$|\varphi \cdot \varphi>$~ to ~$T$~ more than
from ~$|\Fd \cdot \varphi>$~,~ and this gives the frame independent
answer (\ref{trmul}).

It was proposed in \cite{Kanch1} that the mean longitudinal size of
the region, filled by partons with the energy~$\epsilon$, can grow
like $L(E,\epsilon) \sim  E/\epsilon^2$, that gives
$L_z(Y) \sim L(E, \epsilon \sim m) \sim \exp (y)\sim E/m^{2}$.
Then in such a disk (now looking more like a tube)
one can arrange partons with a finite density (in the volume)
and so make an almost ``black" soft object with the low transparency.
This version is not realised in a perturbation theory - for ladder
diagrams one can simply estimate that $L(E,\epsilon)$ is
always  $\sim 1/\epsilon$.
But for a dense parton gas, the growing ``pressure" can push
``additional" partons in the longitudinal direction and lead to
the growth of $L_z(Y)$.~
It is very complicated to convince if this possibility can be
realized in a field theory, partially because we have no soft
field theories in 4 dimensions, and at the same time there is
no supercritical perturbative Pomeron in lower dimensions.
\footnote{ In this connection it is interesting to note that in
the string theory, that is in fact a soft theory on the string scale
$\kappa$, there can exist a close phenomenon.~
It was remarked (\cite{Suss} and later works), that the longitudinal
size of the fast string grows like $\sim E/ \kappa^2$ due to the
inclusion of more and more high frequensity internal harmonics in game.
But the natural question - if we can learn from here something about the
possible structure of the soft \Fd - remains open, because here
we necessary must include a gravitational interaction on the same scale,
and that corresponds to the theory very far from QCD,
and so needs additional considerations (see remarks in section 6).}.

\vspace{1pt}
2) The other way is also to include an increase of the partons density
with $y$,  but now we don't put them in the longitudinally elongated
part of \Fd - rather than force them to increase their mean transverse
momenta and virtuality.
This last one is very natural in a renormalizable field theory like QCD
 - even this behavior shows the solutions of the BFKL equations,
in that the partons spread in transverse momenta with the growth of $y$.
So one can expect that the mean transverse momenta of
partons will grow with $y$. As a result the mean cross-section
of their interaction $\sigma _{0}$ will effectively depend on
$y$ and $Y-y$. This changes the expression (\ref{trimp}) for the
transparency and enlarges the classes of function $\rho (y,b)$ for
that  $T$ is frame independent.
And, like in the previous case, for a very dense \Fd the
~$|\varphi \cdot \varphi>$~ component will dominate in the expression
for~$T$.~
This question will be discussed in the next section.~~
The possibility that the high $k_{\perp}$ can dominate in saturated parton
configurations  was mentioned in an number of papers, starting
from \cite{LRfroi}, and is the most natural way for explaining
various properties of \Fd in QCD including the transparency.

\section*{\bf 3. Partonic structure of a hard  \F disk in QCD. }
\nfn{3}
\vspace{2pt}

In this section we discuss the parton structure of \F~,~ inspired by
various perturbative generalizations of the BFKL approach.
What we firstly need is an approximate qualitative picture that shows
how by growing of $Y$ the \Fd is filled with partons-gluons
at various virtulities (transverse momenta).~
For this purpose we supplement the BFKL equation with terms
describing a gluon fusion \cite{GLR} and also include in it the running
coupling constant.
To simplify considerations we also split, in the kernel of the BFKL
equation (in the manner presented in \cite{jlo}~), the diffusion
processes in $u=\log(k_{\perp})$
and express them in the differential (local in $u$) form,
and the DGLAP processes, for that one can use a trivial
(only the singular part) kernel.
Such a model can be represented by the following equation for the
gluon density $f(y,u)$:
\beal{eveq}
  \frac{\d f(y,u)}{\d y}~~=~~\delta \cdot \alpha_c(u).f(y,u) +
  B_c \alpha_c (u)\cdot   \frac{\d^2 f(y,u)}{\d u^2} +...  \nn \\
  ~~~~~~~~~~~~~~~~~~~~~~~~~~~~~~~~~~~~~~- \lambda_2 e^{-u} \cdot
      \alpha_c^2 (u) \cdot f^2 (y,u) +...  \\
  ~~~~~~~~~~~~~~~~~~~~~~~~~~~~~~+~ \int^u_{\sim 0} du_1
   \alpha_{c}(u_{1)}f(y,u_1)+...   \nn
\eea
The first line in (\ref{eveq}) gives the BFKL like evolution in $y$
and $u$ with the running QCD coupling
~$\alpha _s(u) \simeq  1/(bu + \alpha_0^{-1} )$,~
that is `freezed' on the value ~$\alpha_0$~ in the infrared region.
The second line corresponds to the fusion of partons - where
we have represented (also in the local form) only the first term of
these series, describing the transition of two gluons in
a single one.
The third line represents DGLAP type processes, where a parton can
change fast its $u$ on a small interval of rapidity.
In Eq.(\ref{eveq}) we didn't fix explicitly the coefficients
$(\delta = 12 \log{2}/\pi +\alpha_c~\delta_2~+~...,
~B_c,~\lambda_2,...~)$ -
we only extracted the $\alpha_c$ dependencies from them~
\footnote{
Note, that the choice of gluons as partons is not an unique
one. It is probably more correct to use color dipols \cite{dipole} as
partons - this is more evident and also essential for not meeting the
infrared problems in a Fock wave function.
But, because we are mostly interested in configurations with high
parton numbers, it is probably irrelevant.}.

If we, for a moment, forget about the diffusion of partons over the
virtuality $u$ and also about their gluing, then from Eq.(\ref{eveq})
follows the behavior of the $f(y,u)$ :
\bel{evol1}
f(y,u) ~\sim~  f_g (y,u) =
 f_0 \int_0^y d y_1 ~e^{(y-y_1)\delta_1/u} I_0(2\sqrt{y_1 \xi(u)})
\ee
where $\delta _{1}=\delta /b$~,
~~~$\xi (u) \sim (12/b) \log{(\alpha_c(u_0) / \alpha_c(u) )}$.~~
The evolution, described by Eq.(\ref{evol1}) corresponds to the DGLAP
jump in $u$, followed by the density growth from parton splitting.
At $y > \hat{y}(u) \simeq u^2 \xi(u)/\delta_1^2$  ~(\ref{evol1})
simplifies
\bel{evol11}
 f(y,u) \sim  f_0 (u) ~\exp (~ \delta _1 \frac{y}{u} ~)~~,
\ee
where
$f_0(u) \sim \exp{( \xi(u) /\delta \alpha_c (u) )}$.
At small $y \ll \hat{y}(u)$ the growth of $f(y,u)$ is more close
to the DGLAP type
\bel{bes_ser}
 f_g(y,u) = f_0 \sum_{n=0}^{\infty} \left( \frac{y}{\hat{y}(u)}
\right)^{n/2}  I_n(2\sqrt{ y \xi(u)})  ~\sim~
  \frac{f_0}{(y \xi)^{1/4}} e^{(2\sqrt{y \xi(u)}~)}  ~+~...
\ee
When the density of partons with the virtuality $u$ in some part
of disk becomes very large then the recombination of gluons becomes
essential, and the future growth of $f(y,u)$ with $y$ can stop, and
the density comes to the saturation.
This is described by nonlinear terms in Eq.(\ref{eveq}).
The corresponding limiting value for gluons with the virtuality $u$
can be estimated as
\bel{evol2}
f_{sat}(u) \sim \frac{\delta}{\lambda_2 \alpha_c (u)} \cdot e^{u}
\ee
At these values of $f$ the equilibrium between splitting of
$u$-partons and their joining is reached.
The DGLAP type processes in this region give the small contribution.
In the expression (\ref{evol2}) for $f_{sat}$  only the $\sim f^2$
nonlinear term from (\ref{eveq}) is taken into account.
In fact near the saturation region all nonlinear terms can be
of equal importance.
It probably will only change the coefficient in (\ref{evol2}).
If neglecting the transport of partons in $u$ in Eq.(\ref{eveq}),
their density evolution is described by the equation
\bel{eveqll}
  \d f /\d y ~~=~~  V(f) \equiv   \alpha_c f \cdot ( \delta  -
 \lambda_2 e^{-u} \alpha_c f +
             \lambda_3 e^{-2u} \alpha_c^2 f^2 -...  ~) \nn
\ee
The value $f_{sat}$ corresponds to the point, where
~$V(f_{sat}) = 0$~,~ and because $f$ enters nontrivialy in $V(f)$
only in the combination ~$\alpha_c e^{-u}f$~,~ it will again lead to
(\ref{evol2}).
But if $V(f)$ has no zeros at $f >0$~,~ then there is no complete
saturation.
If $V(f)$ freezes at values $\sim e^u$,~  then, after $f$ reaches
values (\ref{evol2}), a slow universal growth of the parton density
is possible with $y$ like
\be
    f ~\sim~  e^u ~( y - y_0(u) ) ~,
\ee
where $y_0(u)$ can be $\sim u^2$.
The examples of such a $V$ reminding of eiconal or Eq.(\ref{peic}) are
$$
 V(f) ~\sim~ e^u \Big( 1 - \exp (-\lambda \alpha_c e^{-u}~f ) \Big)~~~
 ~~~\hbox{or}~~~
   ~~~~V(f) ~\sim~ (\delta \alpha_c f) /
     \Big( 1 + \lambda \alpha_c e^{-u}~f \Big) ~~~,...
$$
The possibility of such a behavior of $f$ was discussed in a number
of works \cite{jklw}.~
It is hard to understand now what a behavior takes really place in
QCD at high densities, but for our purposes,
connected with the structure of \Fd, it is not so essential.
Probably the best way to approach to this question, and also to the
whole parton structure of the \F limit, is to try using the direct
color field representation for the \Fd, in the manner proposed in
\cite{MLV}, and then to apply the corresponding longitudinal
renorm-groop equations  \cite{jklw}, instead of (\ref{eveq}).
We will not try to do this here, but suppose simply that the full
saturation of the type (\ref{evol2}) takes place.

We need also the approach to the evolution of partons in the
transverse coordinate for a $x_{\bot }$-scale large as compared to
$\Lambda^{-1}_c$.
For that we explicitly include in $f$ the third argument  $x_{\bot }$,
omitted in Eq.(\ref{eveq})
\footnote{
This is the result of simplifications, used in a standart main-log's
approaches to equations of type (\ref{eveq}).
}.
Remember that for the parton evolution, represented by one Regge pole,
the corresponding ``diffusion" equation for density is:
\bel{eveq2}
\d f(y,u,x_{\bot })/\d y ~=~ \Delta (u) \cdot f(y,u,x_{\bot
})+\alpha^{\prime}(u)\cdot \d ^{2} f(y,u,x_{\bot })/\d x_{\bot }^2 ~~,
\ee
Here $\Delta (u)$  is the $u$-dependent intercept, and
$\alpha^{\prime}(u)$ is the $u$-dependent slope of \P representing
the diffusion coefficient (in $x_{\bot }$) for partons with the
virtuality $u$.
The first term at the right hand side of  Eq.(\ref{eveq2}) gives
the parton splitting and corresponds to the same term in
Eq.(\ref{eveq}) with $\Delta (u)=\delta \cdot \alpha_c (u)$.
The second term from r.h.s. of the (\ref{eveq2}) is absent in
(\ref{eveq}), and to combine the two equation we simply add it to
Eq.(\ref{eveq}).
We can get information about the order of $\alpha^{\prime}(u)$ from
the $t$-dependence of \P positions, as it follows from solutions of
the BFKL equation with the running coupling;~ it gives the estimate:
\bel{slop1}
\alpha^{\prime}(u)  \sim  \alpha_c^2 (Q^2) /Q^2
           ~\sim~  e^{-u} \cdot u^{-2}   ~~,
\ee
One point from (\ref{slop1}) is essential for later estimates -
that the slope $\alpha^{\prime}(u)$ rapidly decreases with~$u$.~
As a result we only need the value of $\alpha^{\prime}(u)$
at small(minimal) $u$ to find how fast \Fd expands with $y$.~
Just here the nonperturbative QCD contributions are the maximal ones.
But one can hope that the nonperturbative effect, if correctly
included, will  not change the general structure of equation
(\ref{eveq}) with a frozen $\alpha_s$, as long as we use gluons
as partons.
So we use simply some (phenomenological) value of the
$\alpha^{\prime}$ for the soft $x_{\bot}$ parton diffusion.
Such a parameter enters into the calculation of the velocity of
the expansion of the soft part of the \Fd; and as a result
it follows from (\ref{eveq}, \ref{eveq2}):
\be
R_{soft}(y)=r_0 \cdot y
\ee
This value of $r_0$ fixes the transverse scale in our problem  -
so that later all other quantities can be measured in such $r_0$ units.
Therefore, to simplify all expressions, we will use $r_0$ units
for transverse distances, so always use $r=x_{\bot} /r_0$,
instead of $x_{\bot}$.
Hence at the border of the \Fd we have $r=y$.

All partons belonging to the fast particle fill on average the cone
\FC defined by conditions $(r^2~<~y^2)$ in the $({\bf x}_{\bot},y)$
space. Their section at $y=Y$  gives the soft \Fd, as seen in the
frame where the colliding particle momenta are $\sim e^Y$.
Inside this \FC the soft partons come to the limiting (saturated)
density. Meanwhile we neglect the spreading of the border of this
\FC - we consider it later.

The next question we discuss is: how are the hard partons with the
virtuality $u$ distributed inside the main soft \FC~?
Let's choose some point $(r_1, y_1)$  inside the \FC and consider how
the partons with the virtuality $u $ arise near it.
There are various mechanisms included in equations (\ref{eveq}) that
are responsible for that. Or, in other words, various paths in $(r,y)$
spaces, by that partons evolve to $(r_1, y_1)$ point,
starting their evolution from the top of the \FC.
One can easily show that the main contribution comes from such a path,
illustrated in Fig.3 :
\begin{figure}[h]    \centering
\begin{picture}(150,120)(0,40)
\includegraphics[scale=.3]{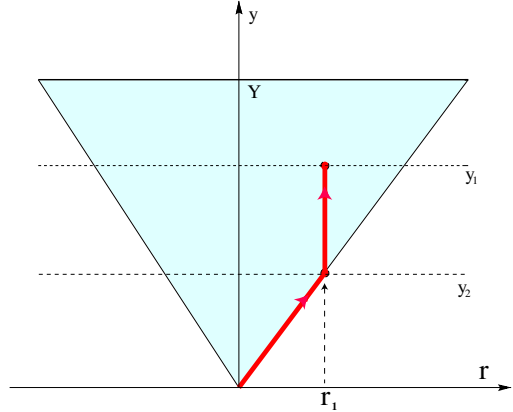}
\end{picture}
\vspace{20mm}
\caption{The main trajectories of parton evolution inside \FC ,
that contributes to the quasiclassical parton dencity. }
\end{figure}
~the partons start to evolve from the soft state with $u \sim 1$
and ``move" not changing their virtuality  from $(r=0,y=0)$
to the point $(r=r_1, y_2=r_1)$, located near the border of the \FC,
then, staying in this point, they jump in $u$  to some higher value
$u_1$, and finally evolve in $y$  at a fixed $r$  to the point
$(r_1, y_1)$, by growing of their density with $y$ like (\ref{evol1}).
  This shows that the mean density of these $u$ partons can be in
two ``states".
If $u$ are large enough and the $u$-partons have had no ``time" to
reach the saturation limit, then their density is defined by the
Eq(\ref{evol1}) with $y \ra y_1 - r_1$.~
From the other side, if $u$-partons had enough long path in $y$ to
reach the corresponding saturation limit, then their density would
be freezed at values given by the Eq.(\ref{evol2}).
These two regimes coincide at the values
\be
u \simeq u_s  = \sqrt{\delta_1 ( y-r )} ~~~~\hbox{ or }~~~~
            r = y - \delta_1^{-1} u^2
\ee
The main asimptotic property of such hard \Fd is that partons for
all virtualities less than $u_s(y,r)$ are in a saturation phase,
and for virtualities greater than $u_s$ the corresponding
parton densities continue to grow with $y$.~
One can combine this all in one approximate expression :
\beal{uden}
f(y,u,r) ~=~ \theta (y -r - \delta_1^{-1} u^2 ) \cdot
f_{sat}(u)~~~+~~~~~~~~~~~~~~~~~~~~~~~~~~~  \nn  \\
~~~~~~~~~ \theta (r - y + \delta_1^{-1} u^2)
\cdot f_g(y,u)
\eea
where $f_{sat}$ is given by (\ref{evol2}), and $f_g(y,u)$ by
(\ref{evol1}). \\
It is interesting to reinterpret this expression  considering
the distribution of parton density at given $u$ as a function
of the transverse distance $r$.
We have accepted, that at the small  $u$ this distribution
is approximately a disk with the radius $R=y$, the soft parton
density inside is $f_{sat}(0) \sim 1$,
and the width of the border in $r$ is also $\sim 1$.
At higher values of $u$ the expression (\ref{uden}) also represents
a $\theta $-like disk - the first term  in the right hand side
corresponds to its interior, and the second to the smeared border.
Such a $u$-disk has a radius $R(u)=y-u^{2}$, the density inside
$\sim f_{sat}(u)$, and the border spread in $r$ by $\sim u/\delta_{1}$,
the form of that is given by the last term in Eq.(\ref{uden}).   \\
\noindent  The second term in Eq.(\ref{uden}) also can be considered
as describing the distribution (over $u$) of parton densities from
the virtualities $u > \sqrt{y}$ up to the values $u \sim y$,
when their densities only grow and are far from a saturation.

Therefore we can represent the parton structure in the \F limit as a
system of enclosed disks, with various virtualities $u$  and radii
$R(u)=Y-u^2$, decreasing with the growth of $u$ and with higher
and higher densities $\sim e^u$~.~
All these $u$-disks are enclosed in the main soft \Fd, and expand with
the same velocity in $y$. The $u$-disks are gray, but their
transparency decreases when $u$ grows.  The total parton density in $r$
is given by the sum of densities of individual $u$ disks ~(see Fig.4),
and can be approximately represented by the enveloping curve in Fig.4~:
\bel{envel}
\rho (r,Y)  \simeq \sum^{\sqrt{\delta_1(Y-r)}}_{u \sim 1}
~f_{sat}(u) ~\theta (Y - \delta_1^{-1} u^2 -r)  ~\sim~
e^{ \sqrt{\delta_1 ( Y-r) } }
\ee
When $Y$ grows then in the middle of the main \Fd a new $u$-disk is
created with the virtuality $\sim \sqrt{y}$  and then it expands like
all other disks with smaller virtualities.
\begin{figure}[h]   \centering
\begin{picture}(150,130)(0,50)
\includegraphics[scale=.3]{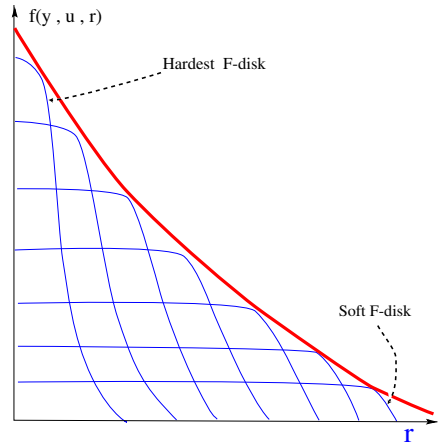}
\end{picture}
\vspace{20mm}
\caption{Tower of  \F-disks with a growing virtualities.
The enveloping curve corresponds to  Eq.(\ref{envel})}
\end{figure}
This picture represents qualitatively the structure of a partonic wave
function of the \F limit, that is in essential part a quasiclassical
one, because the fluctuations of all densities are small compared to
the mean values.~
Evidently such $u$-disks cannot be separated one from another,
and their introduction is used here only to make the presentation
more visual.

At the end of this section let us discuss what are the parameters
of \Fd accompanying fast but very small object like ``onia" with
the size $R_0 \ll \Lambda^{-1}_{QCD}$ ;~ it can be also a virtual
DIS photon that has a size~ $R_0 \sim Q^{-1}$.~
One can expect that the BFKL-like gluon chain is directly attached to
such a small object, with the starting virtuality
$u_R \sim \log R_0^{-2}$.
When $u_R$ is large there is in fact no transverse motion of chain
partons in the $x_{\perp}$ plane. But during the evolution of this
parton chain in rapidity, when going to smaller and smaller momenta,
its density grows like (\ref{evol1}) and finally can reach the
saturation limit corresponding to the virtuality $u_R$.
Also during the BFKL like evolution in rapidity the set of
vitrualities presented in such a gluon(parton) chain expands by the
``diffusion" in $u$.
The virtualities higher than $u_R$ are also reached by the
``DGLAP jumps" in $u$ ( as described by the integral term in
(\ref{eveq}) ) and after that their densities grow as in (\ref{evol1}).
The spreading in $u$ is a such one that the soft virtualities
are reached with a probability of order of unity when
\bel{pomend}
 y  ~>~ y_e(u_R) \sim \alpha_s^{-1} ~(u_R) u_R^2 ~\simeq~ c_e ~u_R^3 ~,
\ee
where ~$c_e  \sim 10^{-2}$ .
~~After that the soft cloud around the $u_0$ particle is created,
and then the soft critical density is reached, the soft \Fd starts
to grow in $\xp$.
Then it produces more hard \F disks, and so on, as described before
for soft ends.
Thus for a small highly virtual object all \F picture is displaced
in $y$ by $c_e ~u^3_R$, so that asymptotically the radius of the soft
\Fd is given by:
\be
R_{soft}(y, u_R)=r_0 ~y  - c_e u^3_R
\ee
More hard \F disks with virtualities $u$ are enclosed in
$R_{soft}$ with the same shift by $c_e u^3_R$ of their radii.

The picture of \F described in this section and based on the
Eq.(\ref{eveq}) corresponds to average (dominant) configurations.
Because such components of \F contain many particles in dence
saturated states, the system is close to a classical one in some
respects, and, as a result, fluctuations around such mean \Fd are
small - they are probably Gaussian ones for a not too large
deviation. \\
\noindent But to treat the transparency of \Fd we need also
probabilities of large fluctuations.
They originate, as in all cascading processes, mainly from a
fluctuation in initial stages of cascading (small $y$), followed
then by the same type of the fluctuations on every step.
In a high energy parton configuration we have $\sim y$ steps
of evolution (parton splitting). On every such a step the density
grows on average. This leads to exponentially damped in $y$
probabilities of rare configurations. At first we need
\bel{sprob}
w_s(y) ~\sim~ \exp (-\gamma_s y)
\ee
where $\gamma_s \sim \Delta$, that gives the probability that for a
fast particle there would be no cascading (or minimal one) at all.
This gives the parton states containing $\sim 1$ soft parton on every
rapidity step and interacting with the target with $Const (y)$
cross-section.

The second probability that can be needed for estimates of $T$
corresponds to configurations, in which we have only  a ``normal" soft
gray \Fd, but hard components of \F are not generated.
Because the hard components can start to grow from every region of soft
\Fd , and at every $y$, we can expect that this probability
\bel{hprob}
w_h(y) ~\sim~ \exp (-\gamma_h y^3) ~,
\ee
where  $\gamma_h \sim \alpha_c (k_{\perp} \sim$~min.~hard~scale~$)$. \\
Let's note that the probability for the creation of a hole with the
area $S_h$ on the hard disk, through that another colliding particle
can penetrate is  $~\sim \exp{(-\gamma_h y S_h)}$~.
This also can be essential only in frames close to the lab.frame
of one of the colliding particles.

\section*{\bf 4. Collision of two hard \F disks.}
\nfn{4}
\vspace{1pt}
Now we consider the collision of two such hard \F disks and estimate
the transparency in various longitudinal systems.
Firstly consider only the parton configurations close to the mean one.
In the $(\vec{\xp},y)$ space the \Fd *\Fd collision with the definite
transverse distance $B$  can be represented by the intersection of two
cones filled with partons (see the Fig.5).

\begin{figure}[h]  \centering
\begin{picture}(160,145)(80,0)
\includegraphics[scale=.7]{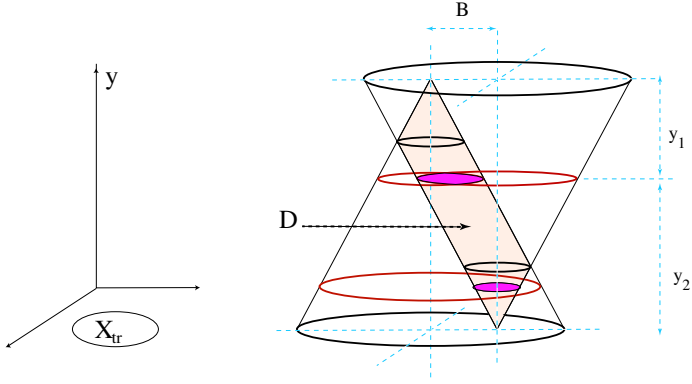}
\end{picture}
\caption{Two intersecting F-cones in $(\xp * y)$ space.
Their sections at rapidities $y_i$ give the picture of two \Fd
collisions in the corresponding frames.
From the region ${\bf D}$ of two \FC intersection particles are
produced.}
\end{figure}
In this picture the choice of the longitudinal Lorentz frame
in that we consider the collision of two \Fd corresponds to
the sections at fixed $y$. As it was discussed in the previous
section, both cones contain embedded subcones with larger
virtualities $u$.  In the process of the collision partons,
that are in the region of \F cones intersection (region $D$ on Fig.5),
interact, and produced secondary particles (jets -for a high
virtuality) fill the region $D$.
Because secondary particles with high $p_{\perp}$ come mainly
from a collision of subcones with a high virtuality - these
particles are concentrated near the line ${\bf L}$ passing
through the central part of $D$.  And therefore at high $Y$
the main contribution to opacity in average configurations
also come from the parton collisions near the line ${\bf L}$.

With the exponential precision the transparency can be expressed
through the parton densities, given by Eq.(\ref{uden}) as:
\be
T ~\sim~ \exp \Big( - \Gamma (Y, y, B) ~\Big) ~,
\ee
where
\bel{gabb}
\Gamma (Y, y, B) ~\sim  \int d^2 \xp \int du_1 du_2 ~
\sigma (u_1,u_2) \cdot f(y, u_1, |{\vec \xp}|)
\cdot f(Y-y, u_2, | {\vec B}- {\vec \xp} |)  ~,
\ee
and
$$
 \sigma (u_1, u_2)  \sim 1/(k_{1\perp} k_{2\perp})
           \sim  \exp(-(u_1+u_2)/2)
$$
is the cross-section for the parton interaction with virtualities
$u_1 , u_2$~.

It is instructive to estimate this integral by choosing two
u-subdisks that are maximally opaque and catch each other during
\Fd collision.  This, in particular, can show the
($(\vec{\xp},y)$)-geometry of main inelastic processes during
the  \Fd * \Fd collision.~
Let us fix some $y$-frame, in that the corresponding rapidities
of $\Fd_1$  and $\Fd_2$  are $y_1$,~ $y_2=Y-y_1$, and the
ipmact parameter $B<Y$.
Next pick out from these ~$\Fd_i$~  the subdisks with such
virtualities $u_1$ and $ u_2$, that these subdisks still overlap,
it is the sum of their radii
\be
r(u_1)+r(u_2) = Y-u_1^2-u_2^2 < B~.
\ee
The area of the overlapping region  ~$S_{12}\sim \int d^2{\vec x}
~\theta~(|{\vec x}| - y_1+u_1^2) ~\theta~(|{\vec x}-B| - y_2+u_2^2)$~
varies with $B$ from $0$, when disks only touch each other, to a
$\min_i~[~\pi~(y_i-u_i)^2~]$~ -  when they overlap completely at $B=0$.
Also, in a such estimate we can don't take into account the spreading
of borders of these u-disks.
Then the simplest quasiclassical estimate of their reciprocal
transparency $T(u_{1},u_{2},B)$ is given by~:
\beal{tran2}
T(u_1,u_2,B) ~\sim~  \exp (~ -\Gamma_{u_1,u_2}
~)~~~~~~~~~~~~~~~~~~  \nn  \\
\Gamma_{u_1,u_2}   ~=~
S_{ab}\cdot \sigma (u_1,u_2) \cdot
f(y_1,u_1,r(u_1))\cdot f(y_2,u_2,r(u_{2}))~.
\eea
In (\ref{tran2}) the most fast changing combination behaves as :
\bel{sff}
\sigma \cdot f\cdot f ~\sim~ f^2_{sat} \exp [(u_1+u_2)/2] ~\sim~
\exp [  (\sqrt{y_1-r_1}+\sqrt{y_2-r_2})/2 ]
\ee
Therefore the maximum  in $\Gamma $ is reached when these $u_i$ disks
touch only each other - that is when $r_1+r_2 \simeq R$.
And this last maximum of $\sigma \cdot f \cdot f$  is reached for
disks with radii:
\be
 r_{1}=\frac{B}{2}+\frac{y_1-y_2}{2};~~~~~
 r_{2}=\frac{B}{2}-\frac{y_1-y_2}{2}
\ee
at $\ |y_1-y_2| < B$ ~~(region $D_2$ on Fig.6)~,~ and the
corresponding virtualities are~:
\bel{u1u2}
u_1=u_2=\sqrt{(Y-B)/2}
\ee
It is correct when disks intersect only partly ~($y$-sections in
region D).~ And when one of u-disks is completely inside the another,
then (Regions $D_1$ and $D_3$):
\begin{eqnarray}
r_1 &=&B,~~~ r_2 \simeq 0,~~~~~ u_{1}=\sqrt{B},~~~~~
u_{2}=\sqrt{y_{1}-B},~~~~~ y_{1} > R+y_{2}    \\
r_1 &=&0,~~~ r_2 \simeq R,~~~~~ u_{2} = \sqrt{B},~~~~~
u_1 = \sqrt{y_2-B} ,~~~~~ y_2 > R+y_1 \nonumber
\end{eqnarray}
We see that the  main contribution to the  $\Gamma$ comes from a
thin region (tube) surrounding the line {\bf L} (see Fig.6) that
passes through the middle of the region $D$.~
\begin{figure}[h]   \centering
\begin{picture}(150,175)(0,40)
\includegraphics[scale=.6]{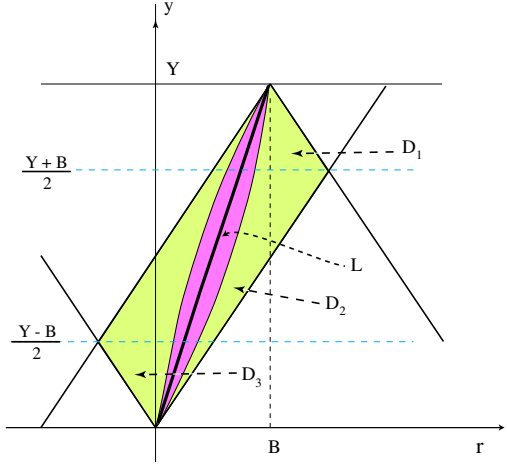}
\end{picture}
\vspace{12mm}
\caption{Two intersecting F-cones.~ Region ${\bf D = D_1+D_2+D_3}$
is where the two \Fd interact.~
In the neighbourhood of the line L the maximum number of partons
(mostly hard) interact.}
\end{figure}
It is the region, in that the maximum number of partons from
the two \Fd collide - and thus from this tube around {\bf L} the
maximum of the secondary particles (gluon jets) is emitted.

Thus to understand if we have boost-invariance of the transparency
$T$ for such mean hard \F configurations we need to consider
the behavior of $\Gamma (u_1, u_2, B)$  along the L line with the
changing $y$.  From (\ref{sff}),(\ref{u1u2}) we see that in the
interval of boosts $-B/2<y<B/2$,  that is when disks not fully
intercept (the section of {\bf L} in the region $D_2$ on Fig.6),
the fastest factor $\sigma f f$ doesn't change with y.
But the factor $S_{ab}$ variates.
On the sections of {\bf L} in  $D_1$ and $D_3$  regions we see the
different behavior of
~$\Gamma \sim \exp \big( \sqrt{y} + \sqrt{Y-B+y} \big)$~~ that is
the growth when we go closer to the edges of \FC along {\bf L} and
becoming maximal on the ends.
And it is opposite to the soft \F picture of previous sections -
there we expected that the transparency is maximal at small $y$ -
that is in the lab.system.
So there is a little hope that the contributions to $T$, coming from
configurations close to the mean one are frame independent.
This is clearly seen for the limiting case ~$B=0$ at
$Y \ra \infty ,~y \ra \infty$, when the integrals in (\ref{gabb})
become more simple. ~Here  the contribution to $T$ can be estimated as
\bel{trh0}
  T(Y,y) ~\sim~  e^{-\sqrt{\gamma_1~ y}} ~+~
        e^{-\sqrt{\gamma_1~ (Y- y)}}  ~,
\ee
where $\gamma_1 = 8\delta_1$.
This expression is simple to understand, because in the exponent
in Eq.(\ref{trh0}) in fact enter the multiplisities of the hardest
saturated subdisks.
It is unclear, if it is a real symptom of conflict with the t-unitarity
or it is enough that only the dominant contribution in $T$
be frame-invariant.

Here, like in the soft eiconal case, the maximum contribution
comes from the rare symmetrical configurations of the
~$|\varphi \cdot \varphi>$~ type, and we end with the frame
invariant answer, when  $T ~\sim~ \exp{( -\gamma Y)}$~.

\vspace{2pt}
\section*{\bf 5. The hard Froissaron in the QCD inspired Regge models.}
\nfn{5}
It would be of some interest to reproduce such a hard partonic
structure of \F in terms of a standard regge approach. This can
be also useful for an unified description of high energy hadronic
data in soft and DIS interactions at the maximal available energies.
It is often believed that to achieve this one can simply put in RFT
the BFKL pomeron with all higher $\alpha_c$ corrections and all
necessary vertices, and supply it with some clever ``soft" boundary
conditions to include effectively the phenomenological \P.
This is probably correct and must directly lead to the desired
results because the equation (\ref{eveq}), from that we started,
reflects directly the BFKL dynamics.
But this way is too general and needs further refinements.

The good starting point can be made from the generalization of
the BFKL equation by the inclusion of the running coupling constant,
discussed in various papers. The main outcome \cite{Lip86} we need
is that in this case \P is represented by the infinite sequence of regge
poles $\P_n$ with intercepts $\Delta_n \sim 1/n$, and that $u_n$ -
the  mean virtualities of $\P_n$~ - grow \cite{jlo} ~like $\sim n$.
The  $\P_n$ components with high $n$ are mainly perturbative, and
the components with small $n$ can have a large nonperturbative
admixture. This especially concerns to the first $\P_1$ component
with a minimal virtuality that can be almost purely
nonperturbative, and so can be directly associated with the
``low-energy" phenomenological \P.

One can hope that such a model, slightly generalizing the usual
one -\P regge approach, can be applied for a better phenomenological
description of various data, also at existing energies;
but it evidently needs additional investigations
\footnote{
One must also slightly ``adjust" some properties of
$\P_n$, that follow directly from BFKL. The residues of such
poles, when calculated from a simple generalisation of BFKL
with a running $\alpha_s$, not aways  give positive vertices.
One can hope that the correct inclusion of higher $\alpha_s$ terms
cure it automatically.}.

The main quantities we need are following. The contribution
of components $\P_n$ of such an ``unitarised" BFKL-like pomeron
can be taken as (\ref{pom1}) in the simplest factorized form
\bel{npol}
 v_n (b,y) ~=~ \frac{g_n \cdot g_n}{\alpha^{\prime}_n y_1} \cdot
~\exp \Big(~ \Delta_n y_1-b^2 / 4\alpha^{\prime}_n y_1 ~\Big) ~~~,
\ee
as corresponding to the generalisation of (\ref{rftl}) to multi -\P
form
\beal{rftl2}
 {\cal L} ~=~
  \sum_n  ~\Big(~ \psi^+_n \frac{~\dlr }{2~\d y}\psi_n  ~+~
  \Delta \psi^+_n \psi_n  ~+~
  \alpha'_n ~\vec{\d_{\perp}} \psi^+_n  ~\vec{\d_{\perp}} \psi_n  ~+~
  (\psi^+_n J_n + J^+_n \psi_n)  ~\Big)  ~+~  \\
  \sum_{m n k} ~r_{m n k} ~(\psi^+_m \psi^+_n \psi_k +
  \psi^+_k \psi_m \psi_n) ~~+ ~...~~~~~~~~~~~~  \nn
\eea
where  : \\
the $n$-th pole intercepts are $\Delta_{n}\sim 1/n$~; \\
the mean  virtualities, corresponding to the n-th pole,
grow like $u_n \sim n$ ; \\
~~$\alpha^{\prime}_n$ - slopes of n-th pole ~$\sim e^{-u_n}\cdot
u_n^{-2}$~~;~  \\
$r_{m n k}$ - 3\P vertices between the corresponding poles;~
now not much can be said about them, but one can expect
\footnote{
If we try to extract the 3\P vertex from gluing 3 BFKL pomerons
with (large) virtualities, we don't find the $\exp (-c u_n)$
behavior, but a large value, coming from the infrared region,
because the corresponding loop integrals are of the type
$\int du \exp(-u)$.
The integrals entering the $r_{m n k}$ vertices are
~$\sim \int du \exp(-u) \chi_m(u) \chi_n(u) \chi_k(u)$~,
~where $\chi_n(u)$ are the internal wave functions of $\P_n$.
The mean $u$ coming from such  $\chi_n(u)$ are large $\sim n$,
but because $\chi_n(u)$ are very flat  the factor $\exp(-u)$
is much more essential, and the dominant contribution is again from
small $u$ and gives $r_{m n k} \sim 1/u_m u_n u_k$.
The above expression for $r_{m n k}$ is in fact a hypothesis -
that when ``all" higher orders in $\alpha_c$ are summed and
also a mixing of the 2-gluon BFKL pomeron with the multigluon
bound states is taken into account, then the resulting
$\P_n$ states $\chi_n(u)$ will be ``sharply" localised in $u$
around the value $\sim n$~.~~
I thank A.B.Kaidalov for many comments and discussions of this
question.
}
that $r_{m n k} \sim \exp (- u_n)$, where $n$ is the maximum
value from $(m,n,k)$,
\\
~$g_n$ - are n-th pole vertices that fast decrease with
$n~$ (~$\sim e^{-u}$)~.

Next we must choose the order of summation of the reggeon diagrams
with $\P_n$.~
If we simply put $\sum v_n$ to the eiconal expression, instead of $v$,
then we become a system of black disks, and it is not a
configuration from that we want to start, as it was explained in
the previous sections.
Perhaps we must firstly take into account the parton recombination,
or in terms of $\P_n$ - their joining described by the vertices~$r_n$.

The corresponding  minimal mechanism  for the reggeon diagrams is
presented by the sum of all tree diagrams. In a quasiclassical
approximation we can neglect the transverse \P  motion inside the
tree diagrams. Because we prepare to use the resulting expressions
mostly inside the \F disks,  where all transverse gradients of the
density are small, the inclusion of the real transverse motion can
only renormalise the values of parameters, resulting from such an
effective zero $\perp$ dimensional reggeon field theory.
We also, for a simplification, take into account only transitions
between the $\P_n$ with the same $n$ in the process of their joining
in a tree.
Then the full contribution from the sum of all tree diagrams can
be represented in the simple form
\bel{Vn}
v_n \ra V_n
         ~\simeq~ \frac{v_n}{1 +  \lambda_n ~v_n } ~~,~~~
   \lambda_n \sim \frac{r_n}{\Delta_n} ~\sim~  e^{-u_n} ~,
\ee
where $r_{n}$ are  proportional to the $3\P_n$ vertices.~~
Then the quantities $V_n$ itself, or their eiconalised combination
\bel{fr0}
 S_1(b,y) ~=~ \sum_k ~c_k~  \Big( -\sum_n V_n(b,y) ~\Big) ^k~~,
~~~~~~F_1(b,y) ~=~ i(1 - S_1)~~,
\ee
can be considered as a first approximation to hard \F with a
saturation
\footnote{
Note that this step is the most uncertain in this program - but
something equivalent anyway must be done.
From one hand one must stop somehow the growth of \P density -
to, at least, use safely reggeon diagrams.
And, on the other hand, this growth is very likely stopped on
\P densities close to the border of the domain of applicability
of reggeon diagrams, or at all outside it.
Note also that close to the \P saturation region many \P
interactions can be essential, and therefore Eq.(\ref{Vn}) for
$V_n$ is only the qualitative one }.

Next, starting from this point, one can construct higher enhanced
and other irreducible diagrams, that transport high
vitruality components to the larger $B$. For the practical
phenomenology one possible way is the generalization of the
approach presented in \cite{KPTer} by the inclusion of the $\P_n$
components with higher virtualities, that is to use directly $V_n$,
or expressions given by Eq.(\ref{fr0}) as a first approximation.

Here we will not enter into details of such a construction,
but restrict ourselves  with one illustration - the structure
of inclusive processes in the central region of rapitity
corresponding to such a \F.~
For usual soft case the main 2\F diagram gives the contribution
to the particle density :
\be
\rho_1 (Y,y) ~\equiv~
  \frac{~d \sigma}{\sigma d y} \sim \frac{y^2 (Y-y)^2}{Y^2} \nn
\ee
and the screening diagrams with \F loops are cancelled. \\
Consider now the inclusive cross-sections at larger virtualities $u$.
It is more convenient and informative to consider them
in the representation with definite impact parameter $B$
between colliding particles and transverse distance $b$
at that inclusive particle (object) with rapidity $y$ and virtuality
$u$ is measured.
\begin{figure}[h]  \centering
\begin{picture}(150,160)(40,30)
\includegraphics[scale=.3]{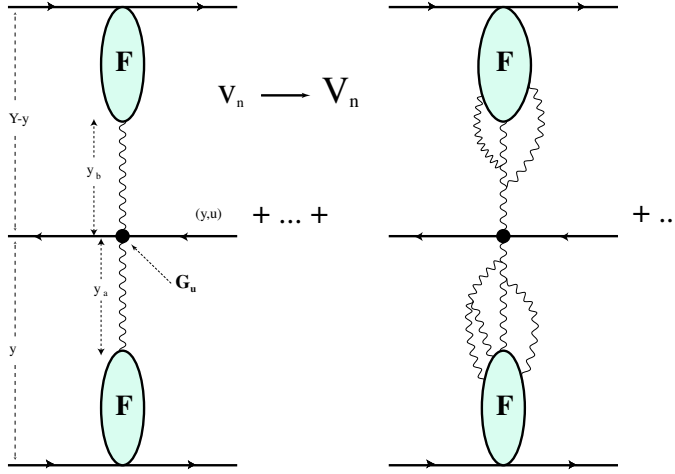}
\end{picture}
\vspace{1.5cm}
\caption{Diagrams for the inclusive production of hard
 objects with a virtuality $u$.
 Wavy lines correspond to $\P_n$ components with $n \sim u$.
 {\bf F} -blocks correspond to the soft $\theta$ -like disks.}
\end{figure}
The main contribution comes from the diagrams Fig.7 :
\bel{incl1}
  \rho_1 (Y,y,B,b,u) ~=~
   F_2 (B-b, Y-y ; u)\cdot {\it G}_n(u)
          \cdot F_2 (b, y ; u )~,
\ee
where the inclusive vertex ${\it G}_n(u) \sim e^{-u} \tilde{g}$~
at $n \sim u$~ and the ``structure function" $F_2$ are  given by
\bel{soed}
   F_2 (b, y ; u ) ~=~ \int dy_a d^2b_a~ F_1(b_a, y-y_a)~ V_n(b-b_a,y_a)
\ee
and the similar expression for other $F_2$.
For large virtualities $u$ the most essential are the $\P_n$ poles
with $n \sim u$.
Also, for a large $n$, the motion in transverse plane is suppressed,
so approximately   ~$v_n(b,y;~ n \sim u) ~\sim~ \delta^2(b)~ v(u,y)$,~
where  ~$v(u,y) \sim \exp{(\Delta_n y)} \sim  \exp{(\delta y/u)}$.~
For $F_1(b,y)$ we take the approximation to a soft gray disk as~~
$f_0
~\theta (ay - b)$ where  $f_0 \sim (1 - T_{soft}$).
This expression for $v(u,y)$ is valid for such an $u$, when the
corresponding partons are far from a saturation,
that is for  $y \ll  u^2/\delta$.~
Taking all this into account we have:
\bel{inc-u }
   F_2 (b, y ; u ) ~=~ \int dy_a F_1~ v(u,y_a)  \sim
     \theta (y - b) \int \limits^{y-b} dy_a v(u,y_a)  ~\sim~
    u~ e^{\delta (y-b) /u} ~\theta (y - b)
\ee
From here, after substituting in Eq.(\ref{incl1}), the simple
expression for the inclusive density follows:
\bel{inc22}
  \rho_1 (Y,y,B,b,u) ~\simeq~
     \tilde{g}(u) ~e^{-u + \delta (Y-b -|B-b|) / u}
           ~\theta (y - b) ~\theta (Y-y - |B-b|)
\ee
It is remarkable that this $\rho_1$ in fact doesn't depend on
$b$  - it is because hard particles are produced from the edge
of the soft disk  $F_1$.

When $y > \eta (u) \simeq  u^2/\delta$~,~ than the
saturation at the $u$ scale begins, and we approximate it by
$v(u,y)~\ra~V=v/(1+\lambda_u v)$~ in $F_2$.~ As a result $F_2$
stops to grow at $v_{max} \sim \lambda^{-1}_u \sim e^u$. \\
Using this substitution we can write the approximate expression,
valid in all the regions of virtualities:
\beal{inc33}
  \rho_1 (Y,y,B,b,u) ~=~
 \frac{ G_u (u)  ~v(u,y-b) ~v(u, Y-y -|B-b|) ~\theta (y - b)
     ~\theta (Y-y - |B-b|) }{  (1 + \lambda (u) v(u,y-b))~
     (1 + \lambda (u) v(u,Y-y-|B-b|)) }   \simeq \nn \\ \nn \\
 ~~~~\simeq  \frac{\tilde{g}(u) ~e^{-u + \delta (Y-b -|B-b|)/u}}
{(1+e^{-u + \delta (y-b)/u})~(1+e^{-u + \delta (Y-y-|B-b|)/u)}}
 ~\theta (y - b)~\theta (Y-y - |B-b|) ~,
\eea
that represents the mean structure of final states produced in
collisions of the enclosed multi-\F disks, considered in the previous
sections.

It is  also not so complicated to write the expression for
$\rho_1$~, in that the soft disk is not $\theta$-like, but more
smooth - as ``normal" amplitude profiles resulting from
several \P exchanges at not ``too asymptotic" energies.

\section*{\bf 6. Brief discussion of some connected questions}
\nfn{6}

We see that QCD leads to a natural picture of \Fd,
that asymptotically becomes more and more black, due to the growth of
the hard parton component with energy. And it is probably the
way how the main requirements from the $t$ and $s$-unitarity can be
fulfilled. \\
In this Section we briefly mention some other interesting aspects of
the \F behavior, that, so or another way, are connected to the
properties of \Fd, considered in the previous sections.

\begin{itemize}
\item
The most often raised question in the connection to \F is if,
at existing energies, we are far away from the region in that
the \F type behavior starts, or may be one can already find some
evident signs of it~?
We don't know this certainly, because there is no common point
of view, what are parameters entering the calculation
of amplitudes.
If the mean number of the \P exchanges $n(s)$  at
$ \sqrt{s} \sim 1 TeV$ is $\simeq 1$ and the multi -\P exchanges
are somehow suppressed, as for example advocated in \cite{land},~
then we must evidently go to much higher energies even to come near
to the \F behavior.
But in most other approaches - like the dual topological string model
(see review \cite{Kaidalov} ), that gives a good description
of a very broad set of data,~ one usually has
$n(\sqrt{s} \sim 540GeV) \sim 2-5$,
and the calculated transparency  is
~$\leq 0.1$ at $B\leq 1 \div 2 GeV^{-1}$.
In such a case we already have in the middle of the fast hadron
a clearly seen embryo of the \F disk.

But even if we are optimistic about a possible fast appearance of
\F, the continuation of such treatments to much higher energies
does not give such large radii of the \F disks, that the regularities
described in the previous section can be clearly seen.
For example, at Planck energies (that are in some sense critical,
because at higher energies new mechanisms can appear) one can estimate
that the \Fd radius
$R_{F}(\sqrt{s} \sim 10^{19}GeV) \sim (20-30)GeV^{-1} \sim 5-6fm$~,
depending on parameters
\footnote{The limiting value of $R_{F}$ alloved by the axiomatic
limitation  is  only $\sim 3 - 5$ times larger, because it is defined
by the $\pi$-meson scale.
But the ``real"  growth of the $R_F$ in QCD  (and in regge
phenomenology) is defined by the larger scale $\sim  0.5 \div 1 GeV$
}.~
Even here the \F radius is close to that of heavy nuclei.
Also the mean transverse momenta of partons in \Fd should be
expected rather small - much smaller than in DIS at already existing
energies. \\
One can also expect, that the events with the final multiplicity
much higher than the mean one are in many respects close in structure
to that one coming from an asymptotic \Fd *\Fd collision.
The estimates of properties of such events from multi-\P exchange
models show also in this direction.~
But, at the same time, we know that the data for such high
multiplicity events don't show the growth of ~$< k_{\perp} >$~.~
So, if the \Fd component already starts to grow, it is so far
only a soft disk.

\item
The heavy nuclei interactions at high (but acceptable on the accelerators)
energies can also have many common points with the collision of \F disks.~
If we consider the interaction at sufficiently high s.c.m. energy
(say 200 GeV per nucleon, like expected in LHC), then the longitudinal
sizes of heavy colliding nuclei become $\sim 1/20 \ll GeV^{-1}$.~
In this configuration the low energy parts of parton clouds from
the individual nucleons fully overlap in the longitudinal directions,
and as a result the density of low momenta partons can essentially
exceed the saturation limit for soft partons.~
Then number of things can happen.
Or all ``superfluous" partons will be simply absorbed and their mean
density approaches to the saturation limit for the soft virtuality scale.
Or the ``additional" partons will be pushed to the region of higher
virtualities, remembering the hard part of \Fd at more higher energies.\\
It should be noted however, that for such a longitudinal joining of
parton clouds we don't need too high energies - it can take place at
energies, already reached in A*A collisions.~
But no growth of the mean $k_{\perp}$ is seen.
Probably it means, that even using this $A^{1/3}$ multiplication of
a wee parton density we do not yet reach the saturation scale even for
soft partons.

\item
At the end let us discuss the question about a possible limitation
from {\em  above} ~on the energies, where \F behavior can be
applied.
And what can be expected for the after-\F  behavior,
if the local field theory is only an approximate one? \\
When we initially are inside the region of applicability of the
local field theory (QCD) and the possible energies are ``in our hands",
then some changes in the \F behavior can be expected only when the
virtualities essential in \F amplitudes
grow with energy and finally reach the limiting flied theory
scale (some physical cutoff connected with the Plank mass,
the string scale,...).
It is possible that in this case any visible changes in \F behavior
will not appear at all, if dominant processes remain somehow soft
at all energies,
or the influence of such small scale processes can be not too
essential.~
For example, in terms of the hard disk parton picture (of Section 3),
it can change some properties of the hard part of the \Fd with the
virtuality $u \sim \log G^{-1}$, and, as a result,
the cross-sections for the particles production at corresponding
$p_{\perp}$.~
But it will not influence the \F behavior of the total inelastic
cross-section and many other cross-sections of soft processes,
because they are controlled by the soft \Fd partons. \\
The other possibility is when some interaction
(like the gravitational one),
now very week, can grow with energy and become dominant, so as to
interfere also with the soft part of QCD interaction.
One can expect that changes in the character of the inelastic scattering
can appear near the Plank scale $\sqrt{s}\sim 10^{19}GeV $~,
where the gravitational interaction starts to be equally important.
The gravitational cross-sections grow like $\sim G^2s$ in the
perturbative region (see for example \cite{AmVen}).
But, as it was proposed in \cite{Banks}, the same character of
the growth of the $\sigma_{in}$ can remain also at the super-planck
energies, where $Gs  \gg 1$~,~ due to the nonperturbative creation
of a large (with a radius $\sim G\sqrt{s}$) events horizon,
in a collision process, that absorbs the colliding particles
and at the end transforms in a black hole.~~
The similar asymptotic behavior of cross-sections is predicted
\cite{Suss} for strings. But it is essential that in these
cases, as opposite to QED and QCD, we can't prepare an analogue of
neutral states to separate somehow the long range elastic and
bremstrahlung part of scattering.  \\
Evidently such questions need an additional investigation to become
more or less clear - but it is not excluded, that after some clever
separation of the elastic component,
and the adequate quantum-mechanical treatment of the event horizon
creation,
and also after imposing $t$ - unitarity
we end again with the same \F-type of the answer for cross-sections. \\
It is interesting to note that in the light of recent propositions
\cite{add} that the real Planck and string scales can be close to the
TeV region, and not to the $10^{19}GeV $ - this question can in principle
become available also to experimental investigations.
In such a case the gravitational contribution to the inelastic
cross-sections can coincide with the hadrons strong interactions
already at $\sqrt{s}\sim 10^3 \div 10^4 GeV$ ~-~ it is close to the
lab.energies $10^{21} \div 10^{23} eV$, at that some strange phenomena
in cosmic rays are discovered.
\end{itemize}

\vspace{10mm}
\nin {\bf ACKNOWLEDGMENTS} \\
\nin I would like to thank K.G.~Boreskov, A.B.~Kaidalov, J.H.~Koch
and K.A.~Ter-Martirosyan for useful conversations and comments.~~
While preparing this paper I often remembered many interesting
discussions with V.N.Gribov about the possible mechanisms,
responsible for Froissart type behavior, that we had had in
1973-1976 years.~~
The financial support of {\bf RFBR} through the grants 98-02-17463
is gratefully   acknowledged.
\vspace{10mm}\\

\end{document}